\documentclass[sigconf]{acmart}

\usepackage{amssymb}
\usepackage{balance}

\usepackage{balance}

\AtBeginDocument{%
  \providecommand\BibTeX{{%
    \normalfont B\kern-0.5em{\scshape i\kern-0.25em }\kern-0.8em\TeX}}}

\begin{document}

\fancyhead{}

\title{It's Food Fight! Introducing the Chef's Hat Card Game  for Affective-Aware HRI.}


\author{Pablo Barros}
\author{Alessandra Sciutti}
\affiliation{\institution{Cognitive Architecture for Collaborative \\Technologies (CONTACT) Unit \\ Istituto Italiano di Tecnologia\\ Genova, Italy}
}

\author{Anne C. Bloem}
\author{Inge M. Hootsmans}
\author{Lena M. Opheij}
\author{Romain H.A. Toebosch}
\affiliation{\institution{Cognitive Architecture for Collaborative \\Technologies (CONTACT) Unit \\ Istituto Italiano di Tecnologia\\ Genova, Italy}
\department{Department of Industrial Design\\
University of Technology Eindhoven \\ Eindhoven, The Netherlands}
}

\author{Emilia Barakova}  
\affiliation{\department{Department of Industrial Design\\
University of Technology Eindhoven \\ Eindhoven, The Netherlands}}

\renewcommand{\shortauthors}{Barros et al.}

\begin{abstract}

Emotional expressions and their changes during an interaction affect heavily how we perceive and behave towards other persons. To design an HRI scenario that makes possible to observe, understand, and model affective  interactions and generate the appropriate responses or initiations of a robot is a very challenging task. In this paper, we report our efforts in designing such a scenario, and to propose a modeling strategy of affective interaction by artificial intelligence deployed in autonomous robots. Overall, we present a novel HRI game scenario that was designed to comply with the specific requirements that will allow us to develop the next wave of affective-aware social robots that provide adequate emotional responses.

\end{abstract}

\keywords{Affective Computing, HRI, Card game }

\begin{teaserfigure}
  \includegraphics[width=\textwidth]{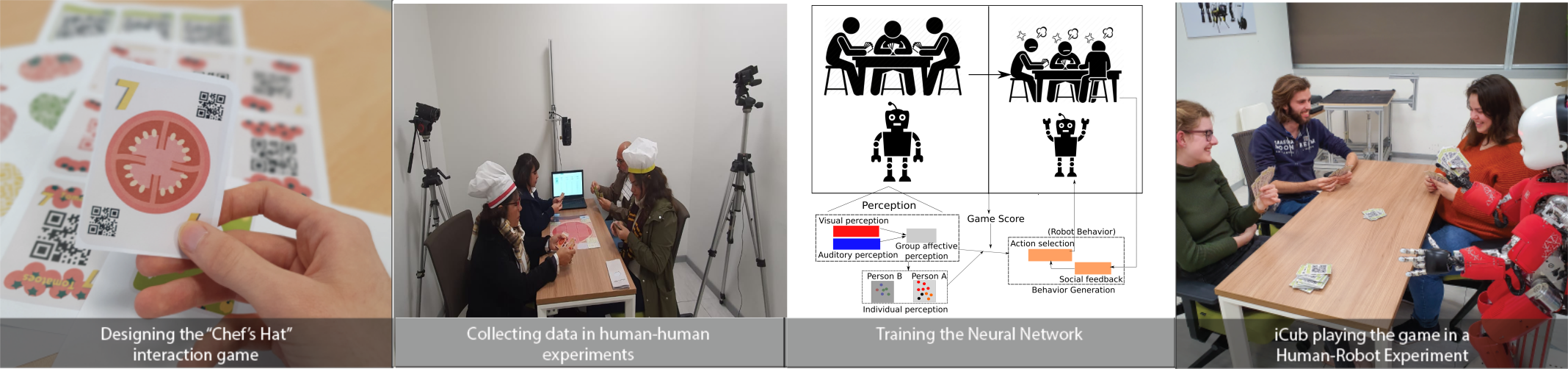}
  \caption{Proposed roadmap for an affective-aware HRI scenario: Developing the Chef’s hat card game, collecting the behavioral measures during the game, designing learning models for autonomous robotics  \cite{barros2019towards}[3], and evaluating the game scenario within human-robot interaction.}
  \label{fig:roadmap}
\end{teaserfigure}

\copyrightyear{2020}
\acmYear{2020}
\acmConference[2020 Workshop Exploring Creative Content in Social Robotics]{2020 Workshop on Exploring Creative Content in Social Robotics at the ACM/IEEE
International Conference on Human-Robot Interaction}{March 23--26,
2020}{Cambridge, United Kingdom}
\acmBooktitle{2020 Workshop on Exploring Creative Content in Social Robotics at the ACM/IEEE
International Conference on Human-Robot Interaction (HRI '20), March 23--26, 2020,
Cambridge, United Kingdom}
\acmDOI{}
\acmISBN{}

\maketitle

\section{Introduction}



Associating perception and action within a continuous adaptation mechanism could improve the way social robots interact and are accepted in real-world social scenarios  \cite{crossman2018influence}. An affective architecture, developed  in the framework of cognitive computation, is the focus of several researchers over the past decade \cite{franklin2013lida,tanevska2018designing}. In particular, modulating behavior based on affect perception has been explored by the affective computing community \cite{hirokawa2018adaptive}. Most of the presented solutions, however, developed social interaction strategies based on emotion contagion \cite{van2018generic} by repeating what was
perceived, or relied upon simple decision trees for generating behavior \cite{tuyen2018emotional}. 

Such solutions are suited for the simple interaction scenarios in which they were deployed. The limitation on designing, implementing and deploying realistic interaction scenarios, where affective information is more complex to capture and respond to is one of the problems that must be solved before the deployment of robots in real life. In a previous attempt without using a robot, a game especially designed to provoke to favor one and disfavor another player in a multiplayer interaction was shown to provoke emotions \cite{barakova2015automatic}, which lasted during the repetitive games between the same players \cite{gorbunov2017memory}.

In this paper, we continue our efforts (refer to our previous work \cite{barros2019towards}) on designing a social interaction scenario, based on a multiple-players games, to be used in affective-aware Human- Robot Interaction (HRI), also illustrated in Figure \ref{fig:roadmap}. The new game, entitled The Chef’s hat, was designed in an iterative process to guarantee that the display of affective behavior was natural and an integral part of the game dynamics. We first explored the social dynamics between human players as part of our game design. The integration of robots into the game, as active players, is of fundamental importance. To facilitate that, we included elements in the games requirements and final mechanics that allow robots to have an estimation of the game status and the person’s social behavior, and play without disturbing the game flow. As the game is intended to enforce teaming between the players, the occurrence of human- robot teaming behavior is an important part of our game design decisions.
The proposed game scenario provides a platform where different learning systems for affective perception, decision making, and behavior generation can be integrated, deployed and evaluated properly and in a close-to-real-world scenario.


\section{The Chef's Hat Interaction Game}

The design of the card-game was based on two main principles:
1) To provide restricted, but natural interaction, where affective behavior naturally arises; 2) to make possible that the game is played between people and a robot and provide a turn-taking opportunities, where a robot, such as \cite{metta2008icub}, has a supportive infrastructure and capacity to process incoming information and generate behavior without breaking the fluidity of the interaction.

\subsection{Interaction Requirements and Initial Assessments}

To maintain the two main principles, as listed in the previous subsection, the card game was design to satisfy the following requirements:

\begin{itemize}
  \item \textbf{R1} - The game should elicit a multitude of affective behavior that the robot can properly understand and model. 
  
   \item \textbf{R2} - The game should be playable without the need of verbal expressions as part of the game mechanics to facilitate the affective understanding and processing.
  
  \item \textbf{R3} - The game should provide the possibility of creating strategies based on affective bonding between the players. This way, interacting with other players through the game should be part of the game’s natural flow.
  
   \item \textbf{R4}  The game should have specific mechanics that, when used, cause affective reactions in the players.

  \item \textbf{R5} - The game should be easy to follow and to understand, having clear turns between the players. The number of actions to be made should be small and easy to process. This way, the limitations of the robot regarding decision-making, behavior display and processing time will be reduced.
 
  \item  \textbf{R6} - The game status and players' behavior should be easy to track and monitor using automatic mechanisms (cameras and microphones). This way we can create a knowledge repository to leverage the learning of gaming strategies and behavior understanding by the robot.
  
    \item \textbf{R7} - The game should give players enough opportunity to interact with all other players through game-mechanics. Actions taken should not only affect the next player in the game, as this could limit both the social and the competitive strategies.

\end{itemize}
Taking these requirements into consideration, we started our design process by evaluating common card games that are well-known to persons from different cultural backgrounds: e.g. UNO, Great Dalmuti and Quartet. Each of these games addresses some of the requirements we listed above, see Table 1. Once these games were selected, we performed several rounds of experiments with four participants playing each of these games.

Taking these requirements into consideration, we started our design process by evaluating common card games that somehow evoke the requirements that are well-known to persons from different cultural backgrounds: e.g. UNO, Great Dalmuti and Quartet. Each of these games addresses some of the requirements we listed above. Once these games were selected, we performed several rounds of empirical experiments with four players playing each of these games.

We performed a behavioral observation to evaluate which requirements each of these games addressed. We video-recorded and analyzed the behavior of all the players  during the game and the behavior observation was carried out by four industrial design students , which were trained on evaluating user interactions. 

\begin{table}[]
\begin{tabular}{c | c | c | c | c | c | c | c | c}
 Game &  R1 & R2 & R3 & R4 & R5 & R6 & R7 \\ \hline
 UNO            & \checkmark &            &             &  \checkmark          & \checkmark   & \checkmark    &               \\
 Great Dalmutti & \checkmark & \checkmark & \checkmark            &            & \checkmark   & \checkmark    &  \checkmark   \\
 Quartet        & \checkmark &            &     \checkmark        &            &              &               &  \checkmark   \\
 Skipbo         &            & \checkmark &             &            &              & \checkmark    &               \\
 Shithead       &            & \checkmark &             &            & \checkmark   & \checkmark    &               \\
\end{tabular}
\caption{Summary of which requirements are addressed by each of the tested games.}
  \label{tab:Requirements}
\end{table}

The analysis showed that popular games, such as UNO and Great Dalmutti, fit most of the defined requirements, but not all. In particular, as discussed in our previous work \cite{barros2019towards}, the possibility of causing adverse affective reactions throughout the gameplay was very present in UNO, but not in the other games. The disadvantage of UNO game was, the use of many words, and the lack of play elements that encourage bonding throughout the entire gameplay. Therefore,  UNO was also disregarded as an appropriate game for the defined purposes.

Great Dalmuti, which is by itself based on a public domain game named Presidenten, or Daifugo in Japan, and UNO were taken as inspiration for creating the new card game that will fit the experimental purpose. The new game presents very simple game mechanics which is fairly easy to be understood and followed, similar to UNO, but without limiting strategic options as it is composed of different rounds with a continuing score. Furthermore, the hierarchy in the gameplay creates interesting group dynamics as well as opportunities for both competitive and cooperative play.




\subsection{Game Mechanics}

Chef‘s hat is a card game that Our game simulates a kitchen, and it has a role-based hierarchy: each player can either be a Chef, a Sous-Chef, a Waiter, or a Dishwasher. The players try to be  the first to get rid of their ingredient cards and become the Chef. This happens for multiple rounds (or Shifts) until the first player to reach 15 Chef . During every Shift there are three phases:

\begin{itemize}
  \item Start of the Shift
  \item Making Pizzas
  \item End of the Shift
\end{itemize}

At the start of the Shift, the cards are divided. Then, the Dishwasher has to give the two cards with highest values to the Chef, who in return gives back two cards of their liking. The Waiter has to give their lowest- valued card to the Sous-Chef, who in return gives one card of their liking. If a player has two Jokers at the start of the Shift, they can choose to play their special action: in case of the Dishwasher this is "Food Fight" (the hierarchy is inverted), in case of the other roles it is "Dinner is served" (there will be no card exchange during that the Shift).

Then, the making of the pizzas starts. The person who possesses a Golden 11 card may start making the first pizza of the Shift. A pizza is prepared when ingredient cards are played on the pizza base on the playing field. A pizza is done when no one can (or wants to) put on any ingredients anymore. The rarest cards have the lowest numbers. A player can play cards by laying down their ingredient cards on the pizza base. To play cards, they need to be rarer (i.e. lowest face values) than the previously played cards. The ingredients are played from highest to the lowest number, so from 11 to 1. Players can play multiple copies of an ingredient at once, but always have to play an equal or greater amount of copies than the previous player did. If a player cannot (or does not want) to play, they pass until the next pizza starts.

At the end of the Shift, the new roles are distributed among the players according to the order of finishing, and every player gets the number of points related to their role.

Several pilot data collections have been conducted and videos recorded and the qualitative observation of these data showed that the requirements indicated before have been met. More specifically, we observed that the players exhibited clear emotional disruptions when the special actions were invoked. Also, we observed that some players developed group-based strategies against a player which had a higher score. Finally, the addition of the physical attributes was noted to give the players more motivations towards winning the game, which improved greatly their social engagement.

\subsection{Game Elements}

 We started the design process of the “Chef’s Hat”  card game taking Great Dalmuti as inspiration. In order to reduce the feeling of a controlled experiment and make the game feel more like a commercially available card game, a theme was chosen and incorporated throughout all game elements. We opted for a Kitchen theme, where players compete to become the chef of a pizza restaurant. This theme has also been used to clarify and give reasoning to certain rules, as a means to make the game even easier to follow.

Next to this, an important part of the design of the game elements was dictated by the game-state monitoring problem. The goal was to only have the need for one top-view camera to monitor the entire game-state.

\begin{figure}
    \centering
    \includegraphics[width=1.0\columnwidth]{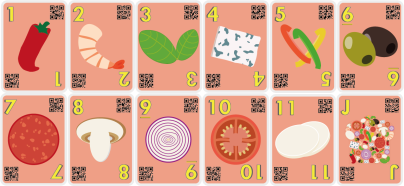}
    \caption{Ingredient cards, and the Joker, with their corresponding face number. The lower the number, the rarer the card.}
    \label{fig:cards}
\end{figure}

The ingredient cards, illustrated in Figure \ref{fig:cards}, needed to be easily recognizable both when played on the playing field and when exchanged among players at the start of the Shift. For this, QR- codes were chosen, as they are hard for humans to recognize or memorize, but easy for the robot or intelligent camera  to read.

\begin{figure}
    \centering
    \includegraphics[width=1.0\columnwidth]{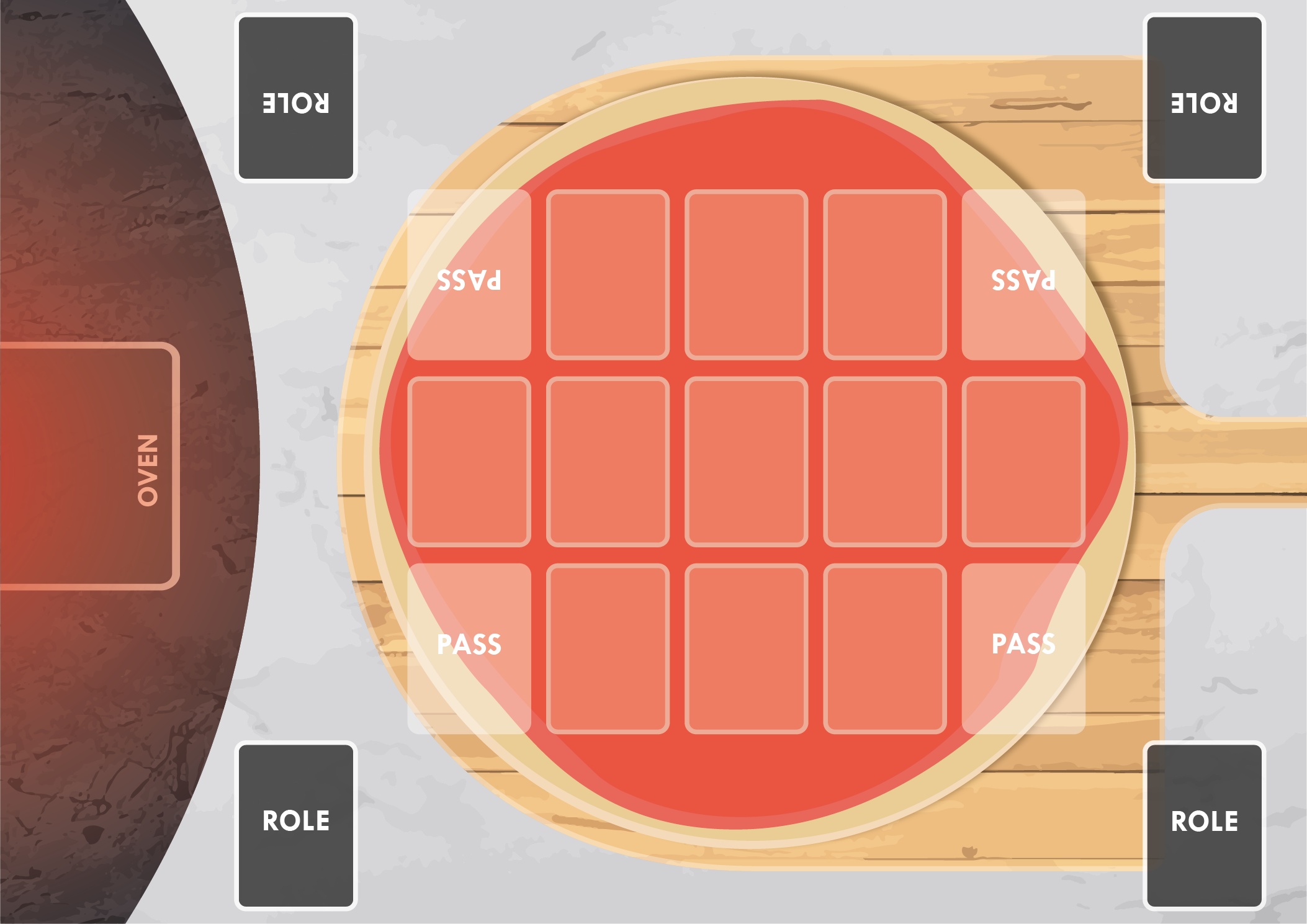}
    \caption{Playing field where the cards are placed, representing a pizza board.}
    \label{fig:field}
\end{figure}

The cards are to be placed on the playing field, illustrated in Figure \ref{fig:field}. While testing the game, a recurrent problem was that players did not play their cards one by one (which made it difficult for the camera to recognize what was played). To counter this, we redesigned the playing field to have 11 different marked places in which players could place their cards on the pizza.

\begin{figure}
    \centering
    \includegraphics[width=0.3\textwidth]{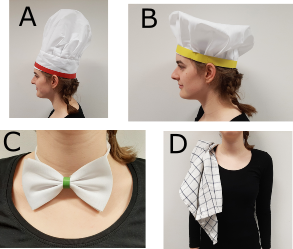}
    \caption{Physical attributes used for each role: A - Chef, B - Sous-Chef, C - Waiter, D - Dish washer.}

\label{fig:attributes}
\end{figure}

Role cards were designed to make it easy to see which player has a certain role. The role cards have an additional functionality to work as a special action card when turned over. To increase the engagement (and possibly competitiveness), it was opted to also have physical role-related attributes: a Chef’s hat, a Sous-Chef’s hat, a Waiter’s bow-tie, and a Dishwasher’s cloth. The attributes were color-coded per role  (see figure \ref{fig:attributes}).

\section{Next Steps: Robot Learning Workflow}

The proposed game was designed to be used in an HRI scenario, in particular to 1) evoke natural and clear emotional reactions and social dynamics in the participants; and 2) to enable a robot participating in the game to sense and monitor both the game status and the affective responses of each participant and of the group as a whole. 

The goal for the robot involved in the game will be to understand the affective behavior of the human participants and use this information, together with that of the game status, to select the best strategy to play the game. "Best strategy" in this sense will acquire different meanings. The robot might either aim at winning the game or at selecting the behavior that maximizes a positive affective reaction in the other players.

One of the most challenging problems we expect to face is the mechanical interaction of the robot with the card game. We do not want to reduce the fluidity, or naturalness of the game from the player’s perspective because this would destroy our general ideal of having a very natural scenario. The iCub is a very advanced social robot, but still have limitations when interacting with real objects, in particular, with very thin objects such as cards. In this regard, we plan to adapt the way that the iCub interacts with the game. Our general idea is to have the iCub asking for help any time it needs to receive or withdraw a card, or put on and off a physical attribute at the beginning of the Shift and use the QR codes present in the card to generate an internal representation of which cards the robot has at hand. However, we are still evaluating the best way to make the robot place a card on the board during the game. Our current investigations involve weather by using a projector that will project the card of the robot on the board or using a specific cardholder that the robot can point at when it wants to place a card. 

To achieve the intelligent and adaptive behavior that  connects the learning of  affective perception/display and game strategy, we envision a cognitive framework which will address the problem based on two perspectives: an adaptive affective perception, based on our previous work on artificial neural networks \cite{barros2016developing, barros2017self}, which will use an unsupervised affective memory mechanism to overcome the problems of online learning; And an interactive reinforcement learning strategy based on learning the social and contextual impact of a certain action by measuring the affective responses of the players, avoiding the necessity of active human interaction in the learning loop.
 
\section{Conclusion}

In this paper, we described our vision on how to achieve an affective awareness in Human- robot interaction. In particular, we propose a new card game that encapsulates specific requirements which makes it suit- able for playing with robots. Moreover, the developed game enables research through design approach, since the game is specially designed to be used as a tool for gathering data and improving the emotional interaction between humans and robots.

Our Chef´s Hat card game has specific elements that integrate continual affective understanding into the game mechanics.Also, it was designed in a way to facilitate learning algorithms that can be used to understand and produce affective behavior, and learn game mechanics .

In the future we will focus on the development of the learning mechanisms to enable the robot to autonomously play the game and we will exploration how such a robot impacts the interaction between the players during the game.

\bibliographystyle{ACM-Reference-Format}
\balance
\bibliography{sample-base}


\begin{thebibliography}{12}


\ifx \showCODEN    \undefined \def \showCODEN     #1{\unskip}     \fi
\ifx \showDOI      \undefined \def \showDOI       #1{#1}\fi
\ifx \showISBNx    \undefined \def \showISBNx     #1{\unskip}     \fi
\ifx \showISBNxiii \undefined \def \showISBNxiii  #1{\unskip}     \fi
\ifx \showISSN     \undefined \def \showISSN      #1{\unskip}     \fi
\ifx \showLCCN     \undefined \def \showLCCN      #1{\unskip}     \fi
\ifx \shownote     \undefined \def \shownote      #1{#1}          \fi
\ifx \showarticletitle \undefined \def \showarticletitle #1{#1}   \fi
\ifx \showURL      \undefined \def \showURL       {\relax}        \fi
\providecommand\bibfield[2]{#2}
\providecommand\bibinfo[2]{#2}
\providecommand\natexlab[1]{#1}
\providecommand\showeprint[2][]{arXiv:#2}

\bibitem[\protect\citeauthoryear{Barakova, Gorbunov, and Rauterberg}{Barakova
  et~al\mbox{.}}{2015}]%
        {barakova2015automatic}
\bibfield{author}{\bibinfo{person}{Emilia~I Barakova}, \bibinfo{person}{Roman
  Gorbunov}, {and} \bibinfo{person}{Matthias Rauterberg}.}
  \bibinfo{year}{2015}\natexlab{}.
\newblock \showarticletitle{Automatic interpretation of affective facial
  expressions in the context of interpersonal interaction}.
\newblock \bibinfo{journal}{\emph{IEEE transactions on human-machine systems}}
  \bibinfo{volume}{45}, \bibinfo{number}{4} (\bibinfo{year}{2015}),
  \bibinfo{pages}{409--418}.
\newblock


\bibitem[\protect\citeauthoryear{Barros and Wermter}{Barros and
  Wermter}{2016}]%
        {barros2016developing}
\bibfield{author}{\bibinfo{person}{Pablo Barros} {and} \bibinfo{person}{Stefan
  Wermter}.} \bibinfo{year}{2016}\natexlab{}.
\newblock \showarticletitle{Developing crossmodal expression recognition based
  on a deep neural model}.
\newblock \bibinfo{journal}{\emph{Adaptive behavior}} \bibinfo{volume}{24},
  \bibinfo{number}{5} (\bibinfo{year}{2016}), \bibinfo{pages}{373--396}.
\newblock


\bibitem[\protect\citeauthoryear{Barros and Wermter}{Barros and
  Wermter}{2017}]%
        {barros2017self}
\bibfield{author}{\bibinfo{person}{Pablo Barros} {and} \bibinfo{person}{Stefan
  Wermter}.} \bibinfo{year}{2017}\natexlab{}.
\newblock \showarticletitle{A self-organizing model for affective memory}. In
  \bibinfo{booktitle}{\emph{Neural Networks (IJCNN), 2017 International Joint
  Conference on}}. IEEE, \bibinfo{pages}{31--38}.
\newblock


\bibitem[\protect\citeauthoryear{Barros, Wermter, and Sciutti}{Barros
  et~al\mbox{.}}{2019}]%
        {barros2019towards}
\bibfield{author}{\bibinfo{person}{Pablo Barros}, \bibinfo{person}{Stefan
  Wermter}, {and} \bibinfo{person}{Alessandra Sciutti}.}
  \bibinfo{year}{2019}\natexlab{}.
\newblock \showarticletitle{Towards Learning How to Properly Play UNO with the
  iCub Robot}.
\newblock \bibinfo{journal}{\emph{arXiv preprint arXiv:1908.00744}}
  (\bibinfo{year}{2019}).
\newblock


\bibitem[\protect\citeauthoryear{Crossman, Kazdin, and Kitt}{Crossman
  et~al\mbox{.}}{2018}]%
        {crossman2018influence}
\bibfield{author}{\bibinfo{person}{Molly~K Crossman}, \bibinfo{person}{Alan~E
  Kazdin}, {and} \bibinfo{person}{Elizabeth~R Kitt}.}
  \bibinfo{year}{2018}\natexlab{}.
\newblock \showarticletitle{The influence of a socially assistive robot on
  mood, anxiety, and arousal in children.}
\newblock \bibinfo{journal}{\emph{Professional Psychology: Research and
  Practice}} \bibinfo{volume}{49}, \bibinfo{number}{1} (\bibinfo{year}{2018}),
  \bibinfo{pages}{48}.
\newblock


\bibitem[\protect\citeauthoryear{Franklin, Madl, D’mello, and
  Snaider}{Franklin et~al\mbox{.}}{2013}]%
        {franklin2013lida}
\bibfield{author}{\bibinfo{person}{Stan Franklin}, \bibinfo{person}{Tamas
  Madl}, \bibinfo{person}{Sidney D’mello}, {and} \bibinfo{person}{Javier
  Snaider}.} \bibinfo{year}{2013}\natexlab{}.
\newblock \showarticletitle{LIDA: A systems-level architecture for cognition,
  emotion, and learning}.
\newblock \bibinfo{journal}{\emph{IEEE Transactions on Autonomous Mental
  Development}} \bibinfo{volume}{6}, \bibinfo{number}{1}
  (\bibinfo{year}{2013}), \bibinfo{pages}{19--41}.
\newblock


\bibitem[\protect\citeauthoryear{Gorbunov, Barakova, and Rauterberg}{Gorbunov
  et~al\mbox{.}}{2017}]%
        {gorbunov2017memory}
\bibfield{author}{\bibinfo{person}{Roman Gorbunov}, \bibinfo{person}{Emilia~I
  Barakova}, {and} \bibinfo{person}{Matthias Rauterberg}.}
  \bibinfo{year}{2017}\natexlab{}.
\newblock \showarticletitle{Memory effect in expressed emotions during long
  term group interactions}. In \bibinfo{booktitle}{\emph{International
  Work-Conference on the Interplay Between Natural and Artificial
  Computation}}. Springer, \bibinfo{pages}{254--264}.
\newblock


\bibitem[\protect\citeauthoryear{Hirokawa, Funahashi, Itoh, and
  Suzuki}{Hirokawa et~al\mbox{.}}{2018}]%
        {hirokawa2018adaptive}
\bibfield{author}{\bibinfo{person}{Masakazu Hirokawa}, \bibinfo{person}{Atsushi
  Funahashi}, \bibinfo{person}{Yasushi Itoh}, {and} \bibinfo{person}{Kenji
  Suzuki}.} \bibinfo{year}{2018}\natexlab{}.
\newblock \showarticletitle{Adaptive behavior acquisition of a robot based on
  affective feedback and improvised teleoperation}.
\newblock \bibinfo{journal}{\emph{IEEE Transactions on Cognitive and
  Developmental Systems}} (\bibinfo{year}{2018}).
\newblock


\bibitem[\protect\citeauthoryear{Metta, Sandini, Vernon, Natale, and
  Nori}{Metta et~al\mbox{.}}{2008}]%
        {metta2008icub}
\bibfield{author}{\bibinfo{person}{Giorgio Metta}, \bibinfo{person}{Giulio
  Sandini}, \bibinfo{person}{David Vernon}, \bibinfo{person}{Lorenzo Natale},
  {and} \bibinfo{person}{Francesco Nori}.} \bibinfo{year}{2008}\natexlab{}.
\newblock \showarticletitle{The iCub humanoid robot: an open platform for
  research in embodied cognition}. In \bibinfo{booktitle}{\emph{Proceedings of
  the 8th workshop on performance metrics for intelligent systems}}. ACM,
  \bibinfo{pages}{50--56}.
\newblock


\bibitem[\protect\citeauthoryear{Tanevska, Rea, Sandini, and Sciutti}{Tanevska
  et~al\mbox{.}}{2018}]%
        {tanevska2018designing}
\bibfield{author}{\bibinfo{person}{Ana Tanevska}, \bibinfo{person}{Francesco
  Rea}, \bibinfo{person}{Giulio Sandini}, {and} \bibinfo{person}{Alessandra
  Sciutti}.} \bibinfo{year}{2018}\natexlab{}.
\newblock \showarticletitle{Designing an Affective Cognitive Architecture for
  Human-Humanoid Interaction}. In \bibinfo{booktitle}{\emph{Companion of the
  2018 ACM/IEEE International Conference on Human-Robot Interaction}}. ACM,
  \bibinfo{pages}{253--254}.
\newblock


\bibitem[\protect\citeauthoryear{Tuyen, Jeong, and Chong}{Tuyen
  et~al\mbox{.}}{2018}]%
        {tuyen2018emotional}
\bibfield{author}{\bibinfo{person}{Nguyen Tan~Viet Tuyen},
  \bibinfo{person}{Sungmoon Jeong}, {and} \bibinfo{person}{Nak~Young Chong}.}
  \bibinfo{year}{2018}\natexlab{}.
\newblock \showarticletitle{Emotional Bodily Expressions for Culturally
  Competent Robots through Long Term Human-Robot Interaction}. In
  \bibinfo{booktitle}{\emph{2018 IEEE/RSJ International Conference on
  Intelligent Robots and Systems (IROS)}}. IEEE, \bibinfo{pages}{2008--2013}.
\newblock


\bibitem[\protect\citeauthoryear{Van~de Perre, Cao, De~Beir, Esteban, Lefeber,
  and Vanderborght}{Van~de Perre et~al\mbox{.}}{2018}]%
        {van2018generic}
\bibfield{author}{\bibinfo{person}{Greet Van~de Perre},
  \bibinfo{person}{Hoang-Long Cao}, \bibinfo{person}{Albert De~Beir},
  \bibinfo{person}{Pablo~G{\'o}mez Esteban}, \bibinfo{person}{Dirk Lefeber},
  {and} \bibinfo{person}{Bram Vanderborght}.} \bibinfo{year}{2018}\natexlab{}.
\newblock \showarticletitle{Generic method for generating blended gestures and
  affective functional behaviors for social robots}.
\newblock \bibinfo{journal}{\emph{Autonomous Robots}} \bibinfo{volume}{42},
  \bibinfo{number}{3} (\bibinfo{year}{2018}), \bibinfo{pages}{569--580}.
\newblock


\end{thebibliography}

\end{document}